\documentclass[a4paper]{article}
\usepackage{amssymb}
\usepackage{amsmath}
\usepackage{overcite}

\newcommand{\ch}{\mathop{\mathrm{ch}}\nolimits}
\newcommand{\sh}{\mathop{\mathrm{sh}}\nolimits}
\renewcommand{\th}{\mathop{\mathrm{th}}\nolimits}

\title{Kink-like Configurations of Interacting Scalar, Electromagnetic, and
Gravitational Fields}

\author{Kulyabov D.\,S., Rybakov Yu.\,P., Shikin G.\,N., Yuschenko L.\,P.
\\
Department of Theoretical Physics, \\
Peoples' Friendship University of Russia,\\
117198 Moscow, RUSSIA, 6, Mikluho-Maklaya str., \\
e-mail: yrybakov@mx.pfu.edu.ru
}

\date{}

\begin{document}

\maketitle

\begin{abstract}
  We have obtained exact kink-like static plane-symmetric solutions to
  the self-consistent system of electromagnetic, scalar, and
  gravitational field equations. It was shown that under certain
  choice of the interaction Lagrangian the solutions are regular and
  have localized energy. The linearized instability of corresponding
  solutions was established both for the case of flat space-time and
  that of interaction with the proper gravitational field.
\end{abstract}

\newpage

\section*{Introduction}
\label{sec:0}

There exist many nonlinear field models, exploiting one
self-interacting field or several interacting fields, that admit the
existence of localized configurations of soliton type. These
configurations are often considered as images of extended elementary
particles \cite{1}. Among such systems the interacting scalar,
electromagnetic, and gravitational fields are the most widely
encountered in nature.

In this paper we consider the massles scalar, electromagnetic, and
gravitational fields for the choice of interaction Lagrangian
explicitly depending on the electromagnetic potentials. For the latter case, after excluding from the system of
equations the scalar field function, we obtain the electromagnetic
field equation with induced nonlinearity containing the electromagnetic
potentials \cite{2}.  Such an interaction appears to be interesting
due to the fact that in the static spherically-symmetric case the
corresponding system of equations admit regular localized solutions of
soliton type both in flat space-time and when taking into account of the
proper gravitational field.  In addition at space infinity
the solutions to the
electromagnetic field equations become those of the linear
electromagnetism. Therefore the dependence on elecromagnetic
potentials emerges only locally, that is in the region of particle
localization where the electromagnetic field seems to be unobservable \cite{3}.
Explicit inclusion of the potentials in the equations of motion is
motivated by the non-existence of regular solutions to the gauge-invariant
nonlinear electromagnetic field equations \cite{4}. We show that
the model suggested admits the existence of kink-like solutions with
localized energy and charge. The configuration of this kind describes
highly polarized system in which positive and negative charges are
localized in the neighbour regions. It will be shown that such a system
is linearly unstable.

\section{Structure of localized solutions}
\label{sec:1}
Lagrangian density for the system of interacting scalar,
electromagnetic, and gravitational fields is choosen as follows:
\begin{equation}
  \label{eq:1}
  L=\frac{R}{2\varkappa} - \frac{1}{4}F_{\alpha\beta} F^{\alpha\beta}
  + \frac{1}{2} \varphi_{,\alpha} \varphi^{,\alpha} \Psi (I),
\end{equation}
where $R$ is the scalar curvature, $\varkappa$ is the Einstein
gravitational constant, $I=A_{\alpha}A^{\alpha}$, $\Psi =1+ \lambda
\Phi(I)$, $\lambda$ is an interaction parameter, $\Phi$ is an arbitrary
function. For $\lambda=0$ we come to the system of scalar and
electromagnetic fields with the minimal coupling.

The static plane-symmetric metric is taken in the form:
\begin{equation}
  \label{eq:2}
  ds^2 = e^{2\gamma(x)}dt^2 -e^{2\alpha(x)}dx^2 -e^{2\beta(x)} \left(
  dy^2 +dz^2 \right),
\end{equation}
where the velocity of light $c=1$, and the functions $\alpha, \beta,
\gamma$ depend only on $x$. In the following we use the harmonic
coordinate condition:
\begin{equation}
  \label{eq:3}
  \alpha=2\beta+\gamma.
\end{equation}

The Einstein's equations for the metric (\ref{eq:2}) under the
condition (\ref{eq:3}) read:
\begin{align}
  \label{eq:4}
  G_0^0 &= e^{-2\alpha} \left( 2\beta'' - 2\gamma'\beta' -
    {\beta'}^2  \right) = -\varkappa T^0_0 , \\
  \label{eq:5}
  G_1^1 &= e^{-2\alpha} \left( 2 \gamma' \beta' +
    {\beta'}^2  \right) = -\varkappa T^1_1 , \\
  \label{eq:6}
  G_2^2 &= e^{-2\alpha} \left( \beta'' + \gamma'' - 2\gamma'\beta' -
    {\beta'}^2  \right) = -\varkappa T^2_2 , \\
  \label{eq:7}
  G^2_2 &= G^3_3, \qquad T^2_2 = T^3_3.
\end{align}

Let us write down the material field equations:
\begin{gather}
  \label{eq:8}
  \frac{1}{\sqrt{-g}} \frac{\partial}{\partial x^\nu} \left( \sqrt{-g}
    g^{\nu\mu} \varphi_{,\mu} \Psi \right) =0, \\
  \label{eq:9}
  \frac{1}{\sqrt{-g}} \frac{\partial}{\partial x^\mu} \left( \sqrt{-g}
    F^{\nu\mu} \right) - \varphi_{,\alpha} \varphi^{,\alpha} \Psi_I
  A^\nu =0,
\end{gather}
where $\Psi_I =d\Psi/dI$.

The energy-momentum tensor of the interacting fields takes the form
\begin{multline}
  \label{eq:10}
  T^\nu_\mu = \varphi_{,\mu} \varphi^{,\nu} \Psi(I) - F_{\mu\alpha}
  F^{\nu\alpha} + \varphi_{,\alpha} \varphi^{,\alpha} \Psi_I
  A^\nu A_\mu - {} \\
  {} - \delta^\nu_\mu \left[ -\frac{1}{4} F_{\alpha\beta}
    F^{\alpha\beta} + \frac{1}{2} \left( \varphi_{,\alpha}
      \varphi^{,\alpha} \right) \Psi(I) \right].
\end{multline}

The field functions depending only on $x$, the electromagnetic field is
determined by the single component of the vector-potential $A_0=A(x)$, being
given by the component of the field tensor $F_{10}=dA/dx$. In this
case $I=A_\alpha A^\alpha = g^{00}A_{0}^2 = e^{-2\gamma}A^2(x)$.

Scalar field equation (\ref{eq:8}) is written in metric (\ref{eq:2})
as follows:
\begin{equation}
  \label{eq:11}
  \frac{d}{dx}(\varphi'\Psi)=0.
\end{equation}

Equation (\ref{eq:11}) has the solution
\begin{equation}
  \label{eq:12}
  \varphi'(x)=C P(I),
\end{equation}
where $P(I)=1/\Psi(I)$, $C$ is the integration constant.

Components of the energy-momentum tensor are:
\begin{gather}
  \label{eq:13}
  T^0_0=\frac{1}{2} e^{-2\alpha} \left[ C^2 P(I) + e^{-2\gamma} {A'}^2
    +
    2C^2 P_I(I)e^{-2\gamma} A^2 \right], \\
  \label{eq:14}
  T^1_1 =-T^2_2 =-T^3_3 =\frac{1}{2}e^{-2\alpha} \left[ C^2 P(I) +
    e^{2\gamma} {A'}^2 \right],
\end{gather}
where $P_I = dP/dI$, $A' = dA/dx$.

Electromagnetic field equation (\ref{eq:9}) in metric (\ref{eq:2}) in
view of (\ref{eq:12}) reads
\begin{equation}
  \label{eq:15}
  \left( e^{-2\gamma} A' \right)' -C^2 P_I e^{-2\gamma} A =0.
\end{equation}

Sum of the Enstein's equations (\ref{eq:5}) and (\ref{eq:6}) with
taking into account of (\ref{eq:14}) leads to the equation
\begin{equation}
  \label{eq:16}
  \beta'' +\gamma'' =0.
\end{equation}

Equation (\ref{eq:16}) has the solution
\begin{equation}
  \label{eq:17}
  \beta(x) =-\gamma(x) +Ex, \qquad E=\textrm{const}.
\end{equation}

Sum of the Einstein's equations (\ref{eq:4}) and (\ref{eq:5}) implies
\begin{equation}
  \label{eq:18}
  \beta'' = -\frac{\varkappa e^{-2\gamma}} {2} \left[ {A'}^2 +C^2 P_I A^2  \right].
\end{equation}

By substituting $A P_I e^{-2\gamma}$ from (\ref{eq:15}) into
(\ref{eq:18}) we get the equation
\begin{equation}
  \label{eq:19}
  \beta''= -\frac{\varkappa}{2} \left( A A' e^{-2\gamma} \right)'.
\end{equation}

We find for (\ref{eq:16}) due to $\gamma(x)$ the analogous equation:
\begin{equation}
  \label{eq:20}
  \gamma''= \frac{\varkappa}{2} \left( A A' e^{-2\gamma} \right)'.
\end{equation}

The first integral for the equation (\ref{eq:20}) is
\begin{equation}
  \label{eq:21}
  \gamma' = \frac{\varkappa}{2} A A' e^{-2\gamma} + k,
\end{equation}
with $k$ being constant.

Let us now consider the special case when $k=0$. Then equation
(\ref{eq:21}) is integrated to:
\begin{equation}
  \label{eq:22}
  e^{2\gamma} = \frac{\varkappa}{2} A^2 + H, \qquad H=\textrm{const}.
\end{equation}

By substituting (\ref{eq:22}) into (\ref{eq:15}) we get:
\begin{equation}
  \label{eq:23}
  \left( \frac{A'}{\varkappa A^2/2 +H} \right)' -C^2 P_I \frac{A} {\varkappa
  A^2 /2 +H} =0.
\end{equation}

Since $dI/dx =2HAA' /(\varkappa A^2/2 + H)^2$, then multiplying
(\ref{eq:23}) by $A'/(\varkappa A^2/2 + H)$, we obtain the equation
which has the first integral
\begin{equation}
  \label{eq:24}
  \left( \frac{A'}{\varkappa A^2/2 +H} \right)^2 - \frac{C^2}{H} P(I) =
  C_1, \qquad C_1 = \textrm{const}.
\end{equation}

Let us consider the solution of the equation (\ref{eq:24}) with
$C_{1}=0$. Then we have finally:
\begin{equation}
  \label{eq:25}
  \int \frac{dA} {\left( \varkappa A^2/2 +H \right)\sqrt{P(I)}} = \pm
  \frac{C (x+x_1)}{\sqrt{H}}, \qquad x_1=\textrm{const}.
\end{equation}

The energy length density of the interacting electromagnetic and scalar
fields appears to be
\begin{multline}
  \label{eq:26}
  E_f = \int T^0_0 \sqrt{-{}^3g}\, dx = {} \\
  {} = \frac{1}{2} \int \left[ C^2 P(I) + e^{-2\gamma} {A'}^2 + 2C^2
    P_I I \right] e^{-\alpha +2\beta} dx.
\end{multline}

After substituting the equality
\begin{equation}
  \label{eq:27}
  e^{-2\gamma} {A'}^2 = \frac{C^2}{H} P(I) e^{2\gamma},
\end{equation}
emerging from (\ref{eq:24}), into (\ref{eq:26})

and using
\begin{equation}
  \label{eq:28}
  dx = \frac{\sqrt{H}}{C} \frac{e^{-2\gamma}}{\sqrt{P}} \,dA,
\end{equation}
which we obtain from (\ref{eq:27}), equality (\ref{eq:26}) becomes
\begin{equation}
  \label{eq:29}
  E_f = \frac{C\sqrt{H}}{2} \int \left[ \sqrt{P} \left( 1+
  \frac{e^{2\gamma}} {H} \right) + \frac{2I P_I}{\sqrt{P}} \right]
  e^{-3\gamma} dA.
\end{equation}

Since $I=e^{-2\gamma}A^2= A^2/\left( \varkappa A^2/2 +H \right)$, then
\begin{equation}
  \label{eq:30}
  e^{-3\gamma} dA = \frac{d\sqrt{I}}{H}.
\end{equation}

Substituting (\ref{eq:30}) into (\ref{eq:29}) we get the following
expression for the field energy length density:
\begin{equation}
  \label{eq:31}
  E_f = \frac{C}{2\sqrt{H}} \int \left[ \sqrt{P} \left( 1+
  \frac{e^{2\gamma}} {H} \right) + \frac{2I P_I}{\sqrt{P}} \right] \,
  d\sqrt{I}.,
\end{equation}
and the invariant energy density:
\begin{equation}
  \label{eq:32}
  T^0_0 \sqrt{-{}^3 g} = \frac{1}{2} \left[ C^2 P(I) + e^{-2\gamma}
  {A'}^2 + 2C^2 P_I I \right]\, e^{-\gamma}.
\end{equation}

Let us find the density of the electric charge $\rho_e$ and the total
charge $Q$. The charge distribution we take from (\ref{eq:9}):
\begin{equation}
  \label{eq:33}
  j^\alpha = -  \varphi_{,\gamma} \varphi^{,\gamma} \Psi_I
  A^\alpha .
\end{equation}

In the case of the static electric field we get from (\ref{eq:33}):
\begin{equation}
  \label{eq:34}
  j^0 =-C^2 e^{-2(\alpha+\gamma)} P_I A.
\end{equation}

Chronometricaly-invariant charge density is defined as
\begin{equation}
  \label{eq:35}
  \rho_e =\frac{j^0}{\sqrt{g^{00}}}  = -C^2
  e^{-2\alpha-\gamma} P_I A .
\end{equation}

The total charge $Q$ is
\begin{equation}
  \label{eq:36}
  Q= \int \rho_e \sqrt{-{}^3 g}\, dx.
\end{equation}

Let us choose the function $P(I)$ as
\begin{equation}
  \label{eq:37}
  P(I)=(1-\lambda I)^2,
\end{equation}
where $\lambda$ is the selfcoupling parameter. Substituting
(\ref{eq:37}) into (\ref{eq:15}) we obtain the equation
\begin{equation}
  \label{eq:38}
  \left( e^{-2\gamma} A' \right)' + 2\lambda C^2 e^{-2\gamma} A -2
  \lambda^2 C^2 e^{-4\gamma} A^3 =0.
\end{equation}

Inserting (\ref{eq:37}) into (\ref{eq:25}), one gets the following
solution of the equation (\ref{eq:38}):
\begin{equation}
  \label{eq:39}
  A(x) = \sqrt{\frac{H}{\lambda -\varkappa/2}} \th bx,
\end{equation}
where $b=\sqrt{C^2(\lambda- \varkappa/2)}$, $-\infty \leqslant x
\leqslant \infty$. The potential $A(x)$ in (\ref{eq:39}) is the regular
function taking finite values
$ -{\varkappa}/{2}$
as $x \to \pm \infty$.

Under substitution of (\ref{eq:39}) into (\ref{eq:32}) we get the
expression for $e^{2\gamma}$:
\begin{equation}
  \label{eq:40}
  e^{2\gamma} = \frac{H\lambda}{\lambda - \varkappa/2} \left( 1-
  \frac{\varkappa} {2\lambda} \frac{1}{\ch^2 bx} \right).
\end{equation}

As follows from (\ref{eq:40}), under the restriction
\begin{equation}
  \label{eq:41}
  H\lambda = \lambda - \frac{\varkappa}{2}>0,
\end{equation}
$e^{2\gamma} \to 1$ as $x \to \pm \infty$. Thus $e^{2\gamma }$ is a
regular function for $x \in [-\infty,\infty]$.

The Einstein equation (\ref{eq:5}) generates, the first integral of
the system of equations (\ref{eq:4}) and (\ref{eq:6}):
\begin{equation}
  \label{eq:42}
  -{\gamma'}^2 + E^2 = - \frac{\varkappa}{2} \left[ -C^2 P(I) +
  e^{-2\gamma} {A'}^2 \right].
\end{equation}
Under sustituting into
(\ref{eq:42}) the function $P(I)$ from (\ref{eq:37}) and also $A(x)$
from (\ref{eq:39}) and $e^{2\gamma}$ from (\ref{eq:40}) we get the
identity if $E=0$. One concludes, in view of (\ref{eq:17}), that
\begin{equation}
  \label{eq:43}
  \beta(x) \equiv -\gamma(x).
\end{equation}

Let us consider the energy distribution, that is the energy density
$T^0_0 \sqrt{-{}^3 g}$:
\begin{multline}
  \label{eq:44}
  T^0_0 \sqrt{-{}^3 g} = {}\\
  {} = \frac{1}{2} \left[C^2 (1- \lambda I)^2 + e^{-2\gamma} {A'}^2 -
    4
    C^2 (1-\lambda I) \lambda I \right]\, e^{-\gamma} = {} \\
  {} = \frac{1}{2} \frac{C^2 (1-\sigma^2)}{\ch^2 bx -\sigma} \left[
    \frac{1-\sigma^2} {\ch^2 bx -\sigma^2} + \frac{1}{\ch^2 bx} -
    \frac{4 \sh^2 bx}{\ch^2 bx -\sigma^2} \right]\, e^{-\gamma},
\end{multline}
where $\sigma^2=\varkappa/2\lambda < 1$.

It follows from (\ref{eq:44}) that the energy density is localized:
\begin{equation}
  \label{eq:45}
  T^0_0 \sqrt{-{}^3 g} \to 0, \qquad \text{при $x \to \pm \infty$}.
\end{equation}

The energy $E_f$ of the interacting $A_\mu$ and $\varphi$ fields is
given by the following expression:
\begin{multline}
  \label{eq:46}
  E_f = \int^{\infty}_{-\infty} T^0_0 \sqrt{-{}^3 g}\, dx = {} \\
  {} = \frac{\lambda C}{2(\varkappa /2)^{3/2}} \left[
    \frac{\sigma}{\sqrt{1 - \sigma^2}} -\frac{\sqrt{1-\sigma^2}}{2}
    \ln \left( \frac{1 + \sigma}{1 - \sigma} \right) \right].
\end{multline}

From (\ref{eq:46}) we deduce that under $0< \sigma <1$, $E_f$ is a
positive bounded quantity.

Let us consider the distribution of charge density $\rho_e$ for the
solution found. From (\ref{eq:35}) we get
\begin{equation}
  \label{eq:47}
  \rho_e = \frac{2 C^2 \sqrt{\lambda} (1 - \sigma^2) \sh bx} {\ch^2 bx
  \sqrt{\ch^2 bx - \sigma^2}}.
\end{equation}

It follows from (\ref{eq:47}) that
$\rho_e = 0$ at $x= 0$, $x= \pm \infty$.

Since $\rho_e$ is an odd function, one finds that the charge $Q$
calculated according to (\ref{eq:36}) vanishes.

\section{Stability analysis}
\label{sec:2}

Let us study the influence of the gravitational field on the
stability of the solutions obtained.
For this purpose we compare
two cases, with the gravitation  being involved or not.

It should be first noticed that the second variation of the total
energy of the system in question turns out to be sign-indefinite
quadratic functional. To check this important property it is
sufficient to consider the case of flat space-time. The modes $A_2$
and $A_3$ being separated one can omit them without loss of generality.

Let us introduce the notations for the perturbations:
\begin{equation*}
  \label{eq:49}
  a_0 := \delta A_0, \quad a_1 := \delta A_1, \quad \xi := \delta \varphi.
\end{equation*}

After tedeons calculations we get for the second variation of the
energy the following expression:
\begin{multline}
  \label{eq:50}
  \delta^2 E = \int_{-\infty}^{\infty}
  \Bigl[ \frac{1}{2} (a_0' - \dot a_1)^2 +
  \frac{1}{2} (\dot \xi^2 + {\xi'}^2)\frac{1}{(1-\lambda A_0^2)^2} -
  {}
  \\
  {} - a_0^2 \frac{\lambda {\varphi'}^2}{(1 - \lambda A_0^2)^4} (1 -
  5\lambda A_0^2) - \frac{\lambda {\varphi'}^2}{(1 - \lambda A_0^2)^3}
  a_1^2 \Bigr]\,dx.
\end{multline}

Sign-alternating character of the functional \eqref{eq:50} makes
very probable the conclusion about the instability of the configuration
in question.
In order to prove
it let us consider the linearized equations for the perturbations. In
particular let us select the equation for the perturbation of the
$A_1$ mode.

\begin{equation}
  \label{eq:51}
  (\dot a_1 -a_0')\dot {} = \frac {2 \lambda {\varphi'}^2}{(1 - \lambda
  A_0^2)^3} a_1.
\end{equation}

If we consider $\Dot a_0'$ as a source then the solution of the
equation (\ref{eq:51}) can be written in the form
\begin{equation*}
  a_1 =a_1^{(0)} + a_1^{(1)},
\end{equation*}
where $a_1^{(1)} $ is a particular solution of the nonhomogeneous
equation and $a_1^{(0)}$ satisfies the homogeneous one. It is
easy to see that $a_1^{(0)}$ increases exponentially with time
thus configuring the instability of the unpertiorbed configuration.

Let us now write down the equation of motion for the $a_1$ with
the gravitation included perturbation
:
\begin{equation}
  \label{eq:52}
  (\dot a_1 - a'_0)\dot {} = \frac{2\lambda {\varphi'}^2}{(1-\lambda
  A_0^2)^3}a_1 - 4 A'_0 \delta \alpha.
\end{equation}

Repeating all the previous arguments
and representing $\dot a'_0 - 4 A'_0 \delta\alpha$ as a
source in the equation \eqref{eq:52},
we confirm the instability of the self-gravitating configuration.

\end{document}